\documentclass[twocolumn,amsmath,amssymb,superscriptaddress,aps,prl,a4paper,floatfix]{revtex4-1}
\usepackage{graphicx}
\usepackage[paperwidth=212mm,paperheight=297mm,centering,hmargin=1.65cm,vmargin=2.6cm]{geometry}
\usepackage{color}
\usepackage{dcolumn}
\usepackage{bm}
\usepackage{physics}
\usepackage{float}
\usepackage{siunitx}
\DeclareSIUnit{\molar}{M}

\begin{document}

\title{Phonon-induced optical dephasing in single organic molecules}

\author{Chloe Clear}
\affiliation{Quantum Engineering Technology Labs, H. H. Wills Physics Laboratory and Department of Electrical and Electronic Engineering, University of Bristol, BS8 1FD, United Kingdom}
\author{Ross C. Schofield}
\affiliation{Centre for Cold Matter, Blackett Laboratory, Imperial College London, Prince Consort Road, SW7 2AZ London, United Kingdom}
\author{Kyle D. Major}
\affiliation{Centre for Cold Matter, Blackett Laboratory, Imperial College London, Prince Consort Road, SW7 2AZ London, United Kingdom}
\author{Jake Iles-Smith}
\affiliation{Department of Physics and Astronomy, University of Sheffield, Sheffield, S3 7RH, United Kingdom}
\author{Alex S. Clark}
\affiliation{Centre for Cold Matter, Blackett Laboratory, Imperial College London, Prince Consort Road, SW7 2AZ London, United Kingdom}
\author{Dara P. S. McCutcheon}
\affiliation{Quantum Engineering Technology Labs, H. H. Wills Physics Laboratory and Department of Electrical and Electronic Engineering, 
University of Bristol, BS8 1FD, United Kingdom}

\date{\today}

\begin{abstract}
Organic molecules have recently gained attention as novel sources of single photons. 
We present a joint experiment--theory analysis of the temperature-dependent emission spectra, zero-phonon linewidth, and second-order correlation function of light emitted from a single molecule. We observe spectra with a zero-phonon-line together with several additional sharp peaks, broad phonon sidebands, and a strongly temperature dependent homogeneous broadening. Our model includes both localised vibrational modes of the molecule and a thermal phonon bath, which we include non-perturbatively, and is able capture all observed features. For resonant driving we measure Rabi oscillations that become increasingly damped with temperature, which our model naturally reproduces. 
Our results constitute an essential characterisation of the photon coherence of these promising molecules, paving the way towards their use in future quantum information applications.
\end{abstract}

\maketitle

Deterministic sources of indistinguishable single photons are a key requirement for many quantum information applications~\cite{ReviewSPS,SinglePhotonQuanCom}. In recent years single molecules of dibenzoterrylene (DBT) have emerged as a promising platform to develop such a source due to a range of desirable properties such as high photostability, high quantum yield~\cite{2004molREV}, favourable absorption and emission wavelengths~\cite{Siyushev2014}, a high branching ratio to the zero-phonon line (ZPL) and 
wavelength tunability across their entire inhomogeneous distribution~\cite{Schadler2019}. This last feature in particular is an advantage over other solid state emitters such as quantum dots (QDs) and diamond colour centres~\cite{inhomoQD,ReviewSPS}, for which it can be difficult to identify two with sufficiently similar emission characteristics on the same sample. DBT can exhibit a lifetime-limited ZPL at cryogenic temperatures $(\leq 4~\mathrm{K})$~\cite{Trebbia09} without any extensive measures to control the local environment such as optical cavities, plasmonic structures or electrical gating. The most promising environment to house DBT molecules is thin nano-crystals of anthracene~\cite{Kyle2015,Polisseni16,Pazzagli2018}. Here they replace three anthracene molecules with little distortion caused to the structure~\cite{Nicolet2007}, and are bonded via van der Waal interactions which helps to reduce environment-induced dephasing~\cite{Grandi2016}. 

With all solid state emitters it is essential that the temperature dependence and nature of the phonon coupling and associated decoherence effects are well characterised and understood. In QDs, for example, the dominant mechanism is coupling of excitons to a bath of longitudinal acoustic phonons, which leads to a broad incoherent sideband in the emission spectrum, as well as broadening of the ZPL at temperatures above $\sim 10~\mathrm{K}$~\cite{iles2016fundamental,Iles-Smith2017,Reigue2017,Sheffield2019Long,muljarov2004dephasing,gold2014two}. These in turn affect the efficiency and indistinguishability of a QD-based single photon source and must be carefully taken into account when designing photonic cavity structures or filtering systems which aim to maximise source figures of merit~\cite{Iles-Smith2017,lodahl2015interfacing}.

In this work we present a detailed experimental interrogation of the optical properties of a DBT--anthracene system, 
and develop a theoretical model which fully captures all observed features, allowing us to uncover the underlying phonon coupling mechanisms. 
The temperature-dependent spectra shown in Fig.~{\ref{OQS}(a-b)} have a rich structure, with a ZPL, several additional narrow lines, and broad sidebands. We are able to associate these with, respectively, direct photon emission, photon emission accompanied by one excitation of a localised vibrational mode of the molecule, and simultaneous emission of a photon and a phonon into the anthracene crystal. 
Closer analysis reveals temperature dependent homogeneous broadening of the ZPL, which in our model arises from 
anharmonicity captured by second order electron-phonon coupling terms in our Hamiltonian. 
These findings have implications for experimental efforts aimed at designing photonic structures to enhance the efficiency and purity of DBT emission \cite{Wang2017,Turschmann2017,Lombardi2017,Grandi2019,Wang2019}. Moreover, the DBT-anthracene crystal is an exemplary open quantum system in its own right, and could be used to test fundamental non-equilibrium concepts such as non-Markovianity.

\begin{figure*}[htp]
\centering
  \includegraphics[width=18cm]{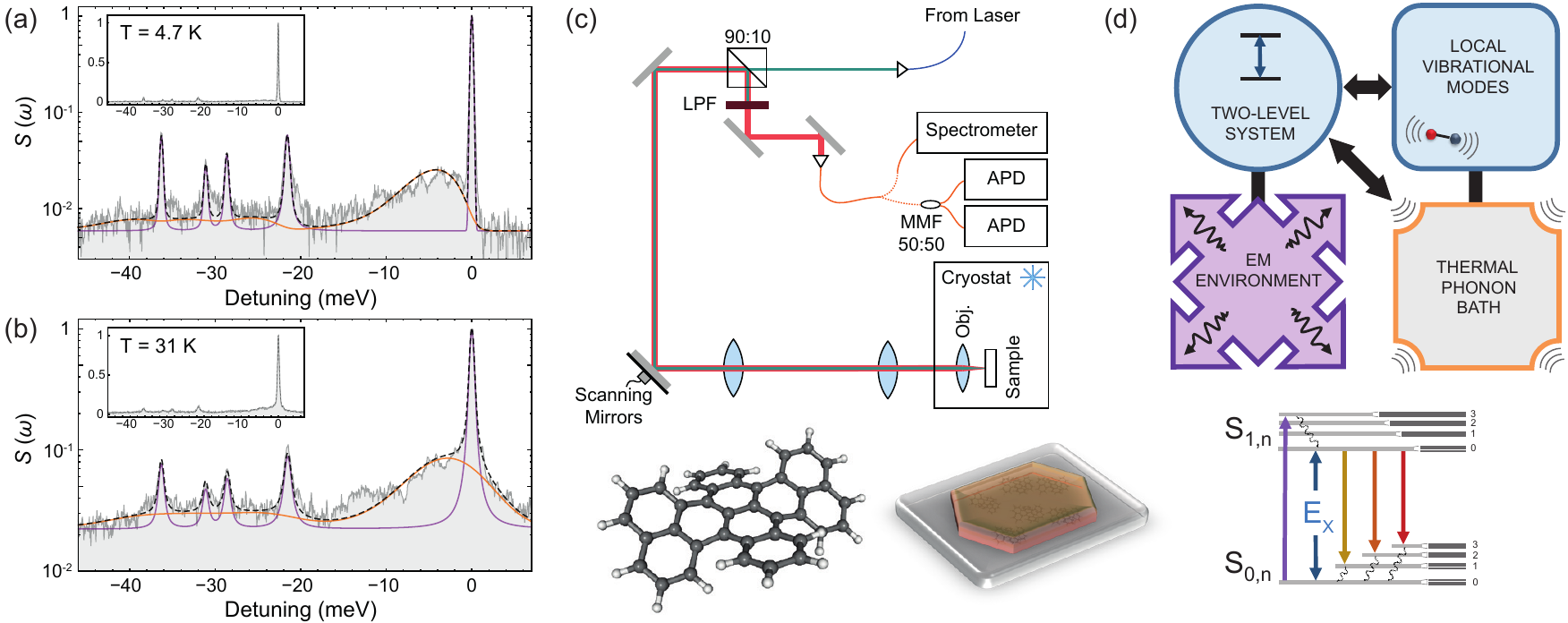}
\caption{Single DBT molecule emission spectra taken at (a) 4.7\,K and (b) 31\,K. Black-dashed lines show the full theoretical model and grey-solid lines show experimental data. The theoretical spectrum showing only the zero-phonon-line and local vibrational mode peaks is shown in purple [c.f. Eq.~{\ref{SZPL}}], while the broad phonon sideband contribution is shown in orange [Eq.~{\ref{SSB}}]. The insets show the spectra on a linear scale. (c) Simplified schematic of the confocal microscope. Dark green is the pump light, and red is the fluorescence. 90:10: 90\% reflection, 10\% transmission beam-splitter; Obj.: Objective lens; LPF: long-pass filter; MMF 50:50: 50\% reflection, 50\% transmission multimode fiber beam splitter; APD: avalanche photodiode. The nano-crystal sample (bottom right) consists of DBT (bottom left) embedded in anthracene. (d) Open quantum system model of a single DBT molecule. The system (blue) contains a two level electronic system (TLS) coupled to a discrete set of vibrational modes and an electromagnetic environment. The thermal phonon bath originates from the nano-crystal and is coupled separately to both system elements. The arrows connected to the TLS represent a non-Markovian interaction including feedback. The schematic energy level diagram shows the ground $S_0$ and excited $S_1$ electronic singlet states with energy splitting $E_X$, and local vibrational modes, all broadened by the thermal phonon environment.}\label{OQS}
\end{figure*}

Our experiments were based on a DBT-doped nano-crystal of anthracene, grown using a re-precipitation technique~\cite{Pazzagli2018}. This was placed in a closed-cycle cryostat incorporated in a confocal microscope shown schematically in Fig.~\ref{OQS}(c). A continuous wave laser was used to excite the DBT molecule to a higher vibrational level of the excited state $S_{1,n>0}$. The molecule then rapidly relaxes to the purely electronic excited state $S_{1,0}$ before decaying to the ground state manifold $S_{0,n}$. The emitted fluorescence was collected by the confocal microscope and dispersed by a grating onto a CCD camera to measure the spectrum. The excitation laser was also tuned over the $S_{0,0} \leftrightarrow S_{1,0}$ ZPL transition for varying illumination intensity while detecting red-shifted photons from the decay of $S_{1,0} \rightarrow S_{0,n>0}$. By splitting this fluorescence on a beam-splitter and monitoring detection times on the two outputs, we measured the second-order correlation function of the emitted light, allowing us to confirm we were dealing with a single DBT molecule. These measurements were then repeated for temperatures from 4.7\,K to 40\,K. A more in-depth description of the experimental methods can be found in the Supplementary Information.

Inspired by the spectra in Fig.~{\ref{OQS}(a-b)} our model of a DBT doped anthracene nano-crystal is shown schematically in Fig.~{\ref{OQS}(d)}. It consists of a two-level-system (TLS) with ground and excited states $\ket{g}$ and $\ket{e}$ split by energy $E_\mathrm{X}$, coupled to the electromagnetic (EM) field, 
harmonic oscillators representing localised vibrational modes of the molecule, and a thermal phonon bath of the anthracene crystal. 
We treat the TLS and localised vibrational modes within our system degrees of freedom, 
and thus capture interactions amongst these to all orders. 
The Hamiltonian of the complete system is
\begin{equation}\label{OGham}
H=H_\mathrm{S}+H_\mathrm{E}+H_\mathrm{I}^\mathrm{EM-TLS}+H_\mathrm{I}^\mathrm{PH-TLS}+H_\mathrm{I}^\mathrm{PH-LV},
\end{equation}
where $H_\mathrm{S}= E_\mathrm{X}\sigma^{\dagger}\sigma+\hbar\sum_{i=1}^N[\Delta_i a^{\dagger}_ia_i +\eta_i \sigma^{\dagger}\sigma(a^{\dagger}_i+a_i)]$, with 
$\sigma=\dyad{g}{e}$. 
The $N$ localised modes 
described by annihilation (creation) operators $a_i$ ($a_i^{\dagger}$) and energy splittings $\Delta_i$ are coupled to the TLS with strengths $\eta_i$. 
The term $H_\mathrm{E}=\hbar\sum_l \nu_l c_l^{\dagger}c_l+\hbar\sum_\mathbf{k} \omega_\mathbf{k} b_\mathbf{k}^{\dagger}b_\mathbf{k}+\hbar\sum_\mathbf{q} z_\mathbf{q} d_\mathbf{q}^{\dagger}d_\mathbf{q}$, contains contributions from harmonic baths describing the EM environment with frequencies $\nu_l$ and annihilation operators $c_l$ for mode $l$, and the thermal phonon baths with frequencies $\omega_\mathbf{k}$ and $z_\mathbf{q}$ and annihilation operators $b_\mathbf{k}$ and $d_\mathbf{q}$ for wavevectors $\mathbf{k}$ and $\mathbf{q}$.

The EM environment--TLS interaction term $H_\mathrm{I}^\mathrm{EM-TLS}$ gives rise to spontaneous emission, 
while  $H_\mathrm{I}^\mathrm{PH-TLS}=H_{\mathrm{I,1}}^\mathrm{PH}+H_{\mathrm{I,2}}^\mathrm{PH}$ couples the thermal phonon bath to the TLS, including terms linear, $H_{\mathrm{I,1}}^\mathrm{PH}=\hbar\sigma^{\dagger}\sigma\sum_\mathbf{k} g_\mathbf{k}(b_\mathbf{k}^{\dagger}+b_\mathbf{k})$, and quadratic $H_{\mathrm{I,2}}^\mathrm{PH}=\hbar\sigma^{\dagger}\sigma\sum_{\mathbf{k}\mathbf{k}'} f_{\mathbf{k}\mathbf{k}'}( b_\mathbf{k}^{\dagger}+b_\mathbf{k})(b_{\mathbf{k}'}^{\dagger}+b_{\mathbf{k}'})$ in the phonon displacements, with coupling constants $g_\mathbf{k}$ and $f_{\mathbf{k}\mathbf{k}'}$ respectively~\cite{FGiustino2017,Non-linMahan}. 
The linear electron-phonon interaction term describes a displacement of the phonon potential well minima. The quadratic term is a consequence of anharmonicity of the thermal phonon modes, resulting in a change of phonon force constants (diagonal) and Raman scattering processes (off-diagonal)~\cite{WiersmaBree1977}. As we will see, the quadratic interaction is crucial for capturing the temperature dependent homogeneous broadening of the ZPL in the emission spectra~\cite{muljarov2004dephasing} 
The final interaction term 
$H_\mathrm{I}^\mathrm{PH-LV}$ 
couples the thermal phonon bath to the localized vibrational modes. 
Full definitions are given in the Supplementary Information.

\begin{figure*}[t!]
\centering
\includegraphics[width=17cm]{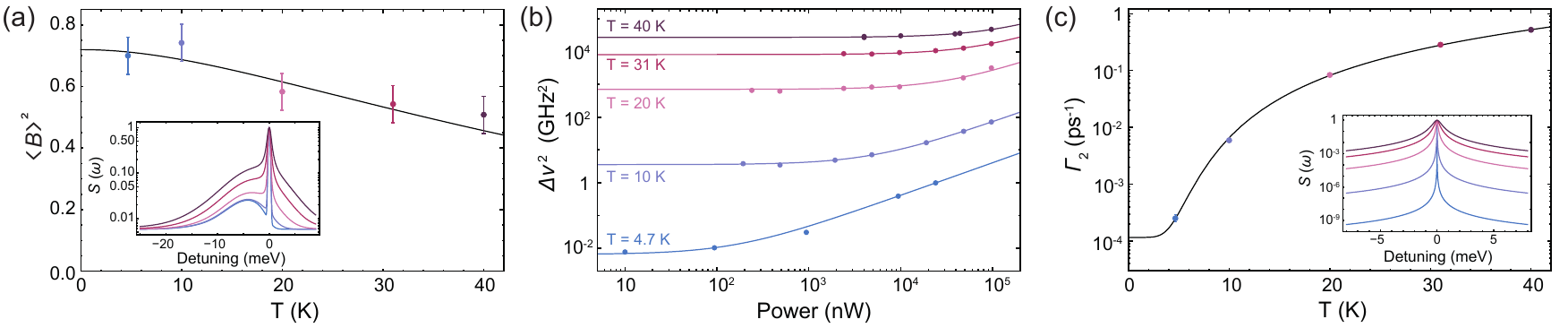}
\caption{ (a) Fraction of emission in the ZPL compared to the broad phonon sideband, not including the local vibrational modes. The solid line shows the theory.   
The inset shows the calculated spectrum of the ZPL and sideband at temperatures where the data was taken. (b) Squared-linewidths extracted from resonant laser scans for varying power at different temperatures. Lines are fits to $\Delta \nu^2=(\Gamma_2 /\pi)^2(1+S)$. (c) Experimental values of $\Gamma_2$ found from the fits in (b), plotted together with prediction from the theoretical model (solid line). The inset shows the calculated ZPL for temperatures at which the data was taken. }
\label{HOM}
\end{figure*}

We now develop a master equation using an extension to the polaron transform approach~\cite{Iles-Smith2017,Nazir2016,1367-2630-12-11-113042,PhysRevB.92.205406}, 
in which we here perform two transformations which displace both the thermal phonon bath and local vibrational modes. 
The first transformation displaces bath phonons dependent on the TLS state, $b_\mathbf{k}\to b_\mathbf{k}+\sigma^{\dagger}\sigma g_{\mathbf{k}}/\omega_{\mathbf{k}}$. This removes the linear TLS--phonon coupling term 
by moving into a basis which includes the distortion of the anthracene lattice in response to the electronic excitation. 
This dresses the TLS with phonon degrees of freedom, which when viewed in the original frame, accounts for non-Markovianity between the TLS and the thermal phonon bath. 
The second transformation acts on the TLS and the localised modes, 
which similarly removes the (linear) interaction terms, and 
dresses the TLS degrees of freedom with those of the local vibration modes. 
We then derive a Born-Markov master equation in the polaron frame~\cite{1367-2630-12-11-113042,Nazir2016,Iles-Smith2017,MasterEqn}, with full details given in the Supplementary Information. 
In a rotating frame and in the Schr\"{o}dinger picture, 
the master equation describing polaron-frame reduced density operator of the TLS and the localised modes is
\begin{align}\label{MEfinal}
&\partial_t \rho(t) =\Gamma_1 \mathcal{L}_{\sigma_a}[\rho(t)]+2\gamma(T) \mathcal{L}_{\sigma^{\dagger}\sigma}[\rho(t)]+\\ 
&\sum_i\Big(\!\!-i\Delta_i [a_i^{\dagger}a_i,\rho(t)]+\Gamma_{i,+}\mathcal{L}_{a^{\dagger}}[\rho(t)]+\Gamma_{i,-}\mathcal{L}_{a}[\rho(t)]\Big),
\nonumber
\end{align}
where $\mathcal{L}_{A}[\rho(t)]=A\rho(t)A^{\dagger}- \frac{1}{2}\qty\big{ A^{\dagger}A,\rho(t)}$, $\Gamma_{i,+}=\kappa_i n(\Delta_i)$ and $\Gamma_{i,-}=\kappa_i( n(\Delta_i)+1)$ with $n(\Delta)=(e^{\hbar\Delta/k_BT}-1)^{-1}$. 
The first term in Eq.~(\ref{MEfinal}) originates from the TLS-EM field interaction and describes 
spontaneous emission with rate $\Gamma_1=1/T_1$ where $T_1$ is the excited state lifetime. 
We note, however, that it contains the dressed dipole operator $\sigma_a=\sigma \prod_i \mathcal{B}_i$ with 
$\mathcal{B}_i=\mathrm{exp}[\eta_i(a^{\dagger}_i-a_i)/\Delta_i]$, and as such accounts for simultaneous emission of a photon and excitation of localised modes. 
The second term describes TLS pure-dephasing with temperature dependent rate $\gamma(T) \propto \sum_{\mathbf{k,k'}}|f_\mathbf{k,k'}|^2 n(\nu_k,T)(n(\nu_k,T)+1) $ which is derived from the quadratic TLS--phonon bath coupling term. 
The local vibrational mode absorption and decay rates $\Gamma_{i,\pm}$ depend on $\kappa$, which is proportional to the local vibrational--phonon bath spectral density. This is taken to be of super-Ohmic form
$J_{\mathrm{PH-LV}}(\Delta)\propto \Delta^3/\zeta^2e^{-\Delta/\zeta}$, where $\zeta$ is the thermal phonon bath cut-off~\cite{Caldeira_LeggettSpecDens1981}.

The emission spectrum is given by 
$S(\omega)=\Re[\int^\infty_{0} d\tau g^{(1)}(\tau)e^{-i\omega\tau}]$ where $g^{(1)}(\tau)=\int^\infty_0 dt \expval{E(t+\tau)^{\dagger}E(t)}$ is the first order correlation function with $E(t)$ the positive frequency component of the electric field operator. Following Refs.~\cite{iles2016fundamental,Iles-Smith2017}, we solve the Heisenberg equations of motion in the polaron frame to find 
$E(t)=E_0(t)+\sqrt{\Gamma_1/2\pi} \sigma_a(t)B_-(t)$, 
where $E_0(t)$ is the free field contribution, assumed to be in the vacuum. We note the second source term contains both TLS and thermal phonon bath degrees of freedom, seen through the appearances of 
$\sigma_a$ and the phonon bath displacement operator 
$B_{\pm}=\mathrm{exp}[\pm \sum_\mathbf{k}g_\mathbf{k}(b_\mathbf{k}^{\dagger}-b_\mathbf{k})/\omega_\mathbf{k}]$. 
We can make use of the varying time scales of the phonon relaxation $(\sim1\,\textrm{ps})$ and photon emission $(\sim1\,\textrm{ns})$ to factorise the correlation function, finding 
$g^{(1)}(\tau)\approx (\Gamma_1/2\pi)g^{(1)}_0(\tau)\mathcal{G}(\tau)$
\label{g1factoring}
where $\mathcal{G}(\tau)=\expval{B}^2 \mathrm{exp}[\phi(\tau)]$, with $\phi(\tau)=\int^{\infty}_0 d\omega J_{\mathrm{PH}}(\omega)\omega^{-2} (\coth(\hbar\beta \omega /2)\cos(\omega \tau)-i \sin(\omega \tau))$ and $\expval{B}=\mathrm{exp}[-\phi(0)/2]$~\cite{Iles-Smith2017}. The electron--phonon spectral density introduced here is
$J_{\mathrm{PH}}(\omega)=\sum_\mathbf{k} g_\mathbf{k}^2\delta (\omega-\omega_\mathbf{k})$, while $g^{(1)}_0(\tau)=\int^{\infty}_0 dt \expval{\sigma_a^{\dagger}(t+\tau)\sigma_a(t)}$. 
We find the emission spectrum can therefore be written 
$S(\omega)\propto S_{\mathrm{ZPL+LV}}(\omega)+S_{\mathrm{SB}}(\omega)$, where 
\begin{equation}
\label{SZPL}
S_{\mathrm{ZPL+LV}}(\omega)=\expval{B}^2\Re\Big[\int^{\infty}_{0}d\tau g^{(1)}_0(\tau)e^{-i\omega\tau}\Big],
\end{equation}
describes peaks associated with the ZPL and localised phonon modes, and 
\begin{equation}
\label{SSB}
S_{\mathrm{SB}}(\omega)=\mathrm{Re}\Big[\int^{\infty}_{0}d\tau g^{(1)}_0(\tau)(\mathcal{G}(\tau)-\expval{B}^2)e^{-i\omega\tau}\Big].
\end{equation}
describes a broad phonon sideband complementing each peak. 
A key advantage of working in the polaron frame is that 
the correlation function $g^{(1)}_0(\tau)$ can be found using the (Markovian) quantum regression theorem~\cite{mccutcheon2015optical,QRT}, while non-Markovian interactions necessary to capture phonon 
sidebands are naturally captured by the phonon bath correlation function $\mathcal{G}(\tau)$ in Eq.~({\ref{SSB}}). Furthermore, 
by writing the spectrum in this way we can immediately see that the Debye--Waller factor (fraction of light not emitted into sidebands) is given by 
$\int S_{\mathrm{ZPL+LV}}(\omega) d\omega/\int S(\omega) d\omega = \expval{B}^2$. 

Predictions of our model are shown by the black dashed curves in Fig.~{\ref{OQS}(a-b)}. The sharp peak at zero detuning 
corresponds to the ZPL at $782.32~\mathrm{nm}$, 
while the other prominent peaks arise from local vibrations of the DBT molecule excited during the photon emission process~\cite{WiersmaBree1977,Grandi2016}. We find that it is necessary to include $N=4$ separate DBT vibrational modes to reproduce these features in the spectra. 
For our model we fit the mode energies $\Delta_i$ and coupling constants $\eta_i$ for 
each temperature and take the averages, resulting in 
$\hbar\Delta_1=(21.55\pm0.01)~\mathrm{meV}$, $\hbar\Delta_2=(28.60 \pm0.01)~\mathrm{meV}$, $\hbar\Delta_3=(31.10\pm0.02)~\mathrm{meV}$ and $\hbar\Delta_4=(36.31\pm0.01)~\mathrm{meV}$, while $\hbar\eta_1=(6.98\pm0.22)~\mathrm{meV}$, $\hbar\eta_2=(6.45\pm0.16)~\mathrm{meV}$, $\hbar\eta_3=(5.73\pm0.09)~\mathrm{meV}$, and $\hbar\eta_4=(9.30\pm0.14)~\mathrm{meV}$. 
To achieve good fits we find it is necessary to include only the ground and first excited state for each vibrational mode in our calculations, meaning that higher vibronic transitions contribute little to the observed spectra. 

The purple curves in Fig.~{\ref{OQS}(a-b)} show the calculated spectra including only the ZPL and local vibrational mode peaks using Eq.~({\ref{SZPL}}), while the orange curves show the phonon sideband contribution given in Eq.~({\ref{SSB}}). The shape of the sideband depends on the functional form of the spectral density $J_{\mathrm{PH}}(\omega)$ which characterises the frequency spectrum of the electron--phonon coupling. 
We use the super Ohmic form 
$J_{\mathrm{PH}}(\omega)=\alpha\,\omega^3 \exp[-\omega^2/\xi^2]$, 
with fitting parameters $\alpha$, which captures the overall coupling strength, and $\xi$ which provides a high-frequency cut-off to reflect the suppression of coupling to phonons whose wavelength is much smaller than the size of the DBT molecule. This form is similar to that used to capture excitation-induced dephasing and phonon sidebands in semiconductor QDs, and can be derived by approximating the electronic ground and excited states as Gaussian wavefunctions~\cite{Nazir2016,Iles-Smith2017,1367-2630-12-11-113042,Sheffield2019Long,Reigue2017,iles2016fundamental}. 

The fraction of the emission which goes into the ZPL and local vibrational mode peaks is given by the Debye--Waller factor, which in our theory is equal to the square of the average phonon bath displacement $\expval{B}^2=\mathrm{exp}[-\int_0^{\infty}J_{\mathrm{PH}}(\omega)\omega^{-2}\coth(\beta\omega/2)d\omega]$. This is plotted as a function of temperature in Fig.~{\ref{HOM}}(a), together with the corresponding experimentally extracted values. 
We see that for this molecule we have a maximum ZPL fraction of 72\%. This is lower than expected for DBT and could partially account for the reduction in coupling observed recently for single molecules in open-access micro-cavities compared to their predictions~\cite{Wang2017,Wang2019}. However, the observed fraction could also be due to the close proximity of surfaces in the nano-crystal host used in these experiments, and further tests with co-sublimation grown crystals~\cite{Kyle2015} may yield a different result. 

Broadening of the emission lines in the spectra is captured by the dissipators in Eq.~(\ref{MEfinal}). 
Of particular interest is the homogeneous broadening of the ZPL with temperature. In our model this broadening follows $\Gamma_2(T)=\Gamma_1/2+\gamma(T)$, where $\gamma(T)$ is a phonon-induced pure dephasing rate. To investigate this broadening in a way that is not affected by the resolution of the spectrometer, we compare our model to measured resonant line scans of the ZPL for varying excitation power. The results at various temperatures are shown in Fig.~\ref{HOM}(b). 
The width of the measured Lorentzian lines can be expressed as $\Delta \nu=\Gamma_2 /\pi \sqrt{1+S}$ with saturation parameter $S$, allowing us to find $\Gamma_2$ by extrapolating the width to zero power, $S\to0$~\cite{Grandi2016}. 
The extracted $\Gamma_2(T)$ are shown in Fig.~\ref{HOM}(c), together with the theoretical prediction. 
The broadening originates from mixing between vibronic states induced by anharmonic effects. This requires the participation of two phonons from the residual bath, and as such necessitates the inclusion of quadratic terms in our Hamiltonian to be captured. Furthermore, the phonon absorption process results in a strong temperature dependence which our model accurately predicts.

\begin{figure}[t!]
\centering
\includegraphics[width=1\columnwidth]{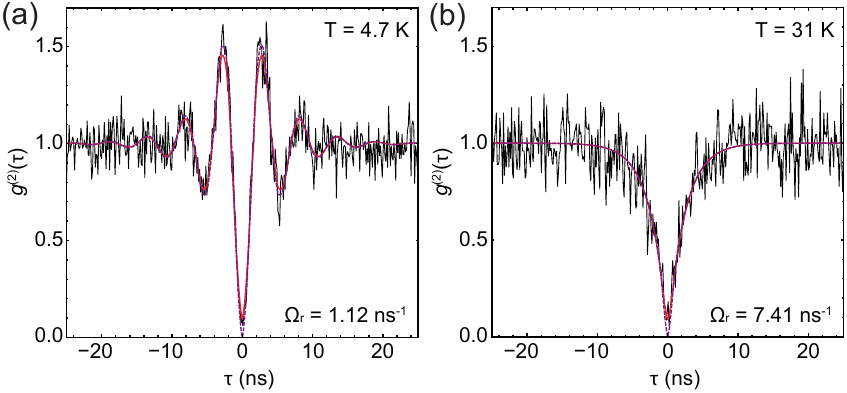}
\caption{Measured $g^{(2)}(\tau)$ taken from the DBT molecule at (a) 4.7\,K and (b) 31\,K. Black shows the experimental data. Red solid lines show the theoretical model convolved with a Gaussian function to account for the detector timing jitter and purple dashed lines show the model without convolution.}
\label{g2fig}
\end{figure} 

To further demonstrate the versatility of our model, we now investigate the time-domain dynamics of the DBT molecule by measuring the second-order intensity correlation function under continuously driven resonant excitation conditions.  To do so we introduce an additional driving term $H_\mathrm{DR}=\frac{\Omega}{2}(\sigma+\sigma^{\dagger})$ to the system Hamiltonian $H_{\mathrm{S}}$ defined in Eq.~({\ref{OGham}}), with Rabi frequency $\Omega$. This results in a slightly modified master equation, the details of which are given in the Supplementary Information. The normalised intensity correlation function is then 
$g^{(2)}(\tau)=\expval{E^{\dagger}E^{\dagger}(\tau)E(\tau)E}_{\mathrm{ss}}/\expval{E^{\dagger}E}_{\mathrm{ss}}^2$, 
where averages are calculated in the steady-state, and $\tau$ is the time delay between detection events \cite{Grandi2016}. 
The calculated $g^{(2)}(\tau)$ and experimental data are shown in Fig.~{\ref{g2fig}}, for temperatures of 4.7\,K in (a) and 31\,K in (b). 
This measurement probes the excited state population of the molecule conditioned on being in the ground state at $\tau=0$. The dip at $\tau=0$ reflects the strong suppression of multi-photon emission events and is characteristic of a single photon source. 
At $T=4.7~\mathrm{K}$ Rabi oscillations can be seen, which represent the coherent exchange of excitations from the driving laser to the system. 
For our calculations we take the molecular parameters extracted from the experimentally measured spectra, with the Rabi frequency $\Omega$ as the only additional fitting parameter. Interestingly, the bare Rabi frequency $\Omega$ that gives the best fit is not the observed Rabi frequency of the oscillations in Fig.~\ref{g2fig}(a). Instead a value of 
$\smash{\Omega_r=\Omega\expval{B}\prod_i \expval{\mathcal{B}_{i}}}$ is observed, which accounts for renormalisation of the bare Rabi frequency arising from phonon coupling~\cite{Dara_2013}. At higher temperatures phonon interactions increasingly damp these oscillations, as is the case in Fig.~\ref{g2fig}(b). 

We have presented a joint experiment--theory analysis that comprehensively 
describes the emission properties of a single DBT molecule encased in an anthracene nano-crystal. The model captures key spectral properties such as the sharp zero-phonon-line, four peaks associated with local vibrational modes of the molecule, and broad phonon sidebands. We also observed a temperature dependent homogeneous broadening of the ZPL, which in our model arises when we include anharmonic effects by taking the electron-phonon interaction to second order in the Hamiltonian. These findings have important consequences for the use of molecules as single photon sources in quantum information applications, as the indistinguishability of emitted photons is strongly affected by the various phonon related features that we identify. 
Our model constitutes a natural starting point for future work investigating effects associated with the coupling of molecules to photonic structures in the form of optical waveguides~\cite{Turschmann2017,Lombardi2017,Grandi2019} and cavities~\cite{Wang2017,Wang2019,PhysRevLett.122.203602}.

\begin{acknowledgements}
We thank Jon Dyne and Dave Pitman for their expert mechanical workshop support. This work was supported by EPSRC (EP/P030130/1, EP/P01058X/1, EP/R044031/1, EP/S023607/1, and EP/L015544/1), the Royal Society (UF160475), and the EraNET Cofund Initiative QuantERA under the European Union’s Horizon 2020 research and innovation programme, Grant No. 731473 (ORQUID Project). J.I.-S. acknowledges support from the Royal Commission for the Exhibition of 1851.

Near the completion of this work we became aware of a similar 
theoretical study investigating the optics of molecular systems encased in crystals~\cite{Reitz2019}.
\end{acknowledgements}

\pagebreak
\widetext
\begin{center}
\textbf{\large Supplementary Information: Phonon-induced optical dephasing in single organic molecules}
\end{center}
\setcounter{equation}{0}
\setcounter{figure}{0}
\setcounter{table}{0}
\setcounter{page}{1}
\makeatletter
\renewcommand{\theequation}{S\arabic{equation}}
\renewcommand{\thefigure}{S\arabic{figure}}
\renewcommand{\bibnumfmt}[1]{[S#1]}
\renewcommand{\citenumfont}[1]{S#1}

\title{Supplementary Information: Phonon-induced optical dephasing in single organic molecules}

In this supplement we detail theoretical and experimental information to support the main text.
\let\newpage\relax
\maketitle

\section{Electron--phonon interaction}

We begin by defining the electron--phonon interaction Hamiltonian as
\begin{equation}\label{Hep}
H_{ep}=\int d^3r\rho(\mathbf{r})V_{ep}(\mathbf{r}),
\end{equation}
where $\rho(\mathbf{r})$ is the electronic charge density of the lattice and $V_{ep}(\mathbf{r})$ is the electron--phonon potential. We expand this potential in powers of small ion displacements $\mathbf{Q}_i$ from the equilibrium position $\mathbf{R_i^{(0)}}$ where $\mathbf{R_i}=\mathbf{R_i^{(0)}}+\mathbf{Q}_i$ such that
\begin{equation}\label{TaylorVep}
\begin{split}
V_{ep}(\mathbf{r})&=-\sum_i\mathbf{Q}_i\cdot\nabla V_{ei}(\mathbf{r}-\mathbf{R_i^{(0)}})+\frac{1}{2}\sum_i\mathbf{Q}_i\cdot\nabla\nabla V_{ei}(\mathbf{r}-\mathbf{R_i^{(0)}})\cdot\mathbf{Q}_i+\mathcal{O}(Q^3),\\
&\approx V_{ep}^{(1)}(\mathbf{r})+V_{p}^{(2)}(\mathbf{r}),
\end{split}
\end{equation}
where $V_{ei}(\mathbf{r}-\mathbf{R_i}^{(0)})$ is the electron--ion potential. The displacements $\mathbf{Q}_i$ can be represented in second quantisation as 
\begin{equation}\label{Qeqn}
\mathbf{Q}_i=i\sum_{\mathbf{k}} \Big(\frac{1}{2NM\omega_\mathbf{k}}\Big)^{1/2}e^{i\mathbf{k}\cdot\mathbf{R_i^{(0)}}}(b_{\mathbf{k}}+b^{\dagger}_{\mathbf{k}})\hat{\xi}_{\mathbf{k}},
\end{equation}
where $b_{\mathbf{k}}$ ($b^{\dagger}_{\mathbf{k}}$) is the phonon annihilation (creation) operator of mode $\mathbf{k}$ with frequency $\omega_{\mathbf{k}}$ and polarisation $\hat{\xi}_{\mathbf{k}}$~\cite{mahan2013many}. Substituting $\mathbf{Q}_i$ into $V_{ep}(\mathbf{r})$ and again into equation (\ref{Hep}) we find the electron--phonon interaction Hamiltonian up to second order \cite{Non-linMahan,Nazir2016}
\begin{equation}\label{Hep1}
H_{ep}^{(1)}=\sum_{\mathbf{k}}\rho(\mathbf{k})M^{(1)}(\mathbf{k})(b_{\mathbf{k}}+b^{\dagger}_{-\mathbf{k}}),
\end{equation}
\begin{equation}
H_{ep}^{(2)}=\frac{1}{2}\sum_{\mathbf{k,k'}}\rho(\mathbf{k+k'})M^{(2)}(\mathbf{k,k'})(b_{\mathbf{k}}+b^{\dagger}_{-\mathbf{k}})(b_{\mathbf{k'}}+b^{\dagger}_{-\mathbf{k'}}),
\end{equation}
with the first and second order matrix elements 
\begin{equation}
M^{(1)}(\mathbf{k})=-\Big(\frac{1}{2NM\omega_\mathbf{k}}\Big)^{1/2}\hat{\xi}_{\mathbf{k}}\cdot \mathbf{k} V_{ei}(\mathbf{k}),
\end{equation}
\begin{equation}\label{M2}
M^{(2)}(\mathbf{k,k'})=\frac{1}{4NM\sqrt{\omega_{\mathbf{k}}\omega_{\mathbf{k'}}}}\hat{\xi}_{\mathbf{k}}\cdot(\mathbf{k+k'})\cdot(\mathbf{k+k'})\cdot\hat{\xi}_{\mathbf{k'}}V_{ei}(\mathbf{k+k'}).
\end{equation}

To calculate dephasing rates in our subsequent master equation we require an analytical form for the linear and quadratic form of the matrix elements. By inserting a resolution of identity and we can re-write the linear and quadratic interaction Hamiltonian as 
\begin{equation}
H_{ep}=\dyad{e}{e}\sum_{\mathbf{k}}g_{\mathbf{k}}(b_{\mathbf{k}}+b^{\dagger}_{-\mathbf{k}})+\frac{1}{2}\dyad{e}{e}\sum_{\mathbf{k,k'}}f_{\mathbf{k,k'}}(b_{\mathbf{k}}+b^{\dagger}_{-\mathbf{k}})(b_{\mathbf{k'}}+b^{\dagger}_{-\mathbf{k'}})
\end{equation}
with the electron--phonon coupling strengths $g_{\mathbf{k}}=\matrixel{e}{M^{(1)}(\mathbf{k})\rho(\mathbf{k})}{e}$ and $f_{\mathbf{k,k'}}=\frac{1}{2}\matrixel{e}{M^{(2)}(\mathbf{k,k'})\rho(\mathbf{k+k'})}{e}$. The off diagonal matrix elements $\matrixel{e}{\dots}{g}, \matrixel{g}{\dots}{e}$ are neglected as phonon energies are not sufficient to drive transitions between the ground and excited state. We have in addition set the ground state matrix element to zero. Substituting the electron density in reciprocal space, $\rho(\mathbf{q})=\int d^3 r \sum_{\lambda,\eta} c_{\lambda}^{\dagger}c_{\eta}\psi_{\lambda}^{\dagger}(\mathbf{r})\psi_{\eta}(\mathbf{r})e^{i\mathbf{q}\cdot\mathbf{r}}$ where $\lambda , \eta =\{e,g\}$, into the linear and quadratic coupling strengths we find 
\begin{equation}
g_{\mathbf{k}}=\sum_{\eta,\lambda}\int d^3r \psi_{\eta}^{\dagger}(\mathbf{r})\psi_{\lambda}(\mathbf{r})e^{i\mathbf{k}\cdot\mathbf{r}} \matrixel{e}{M^{(1)}(\mathbf{k})c_{\eta}^{\dagger}c_{\lambda}}{e},
\end{equation}
\begin{equation}
f_{\mathbf{k,k'}}=\frac{1}{2}\sum_{\eta,\lambda}\int d^3r  \psi_{\eta}^{\dagger}(\mathbf{r})\psi_{\lambda}(\mathbf{r})e^{i(\mathbf{k}+\mathbf{k'})\cdot\mathbf{r}} \matrixel{e}{M^{(2)}(\mathbf{k,k'})c_{\eta}^{\dagger}c_{\lambda}}{e}.
\end{equation}
Substituting in the equation for $M^{(1)}(\mathbf{k})$ and $M(^{(2)}(\mathbf{k,k'})$ and asuming the electron--ion potential is equal to a constant deformation potential such that, $ V_{ei}(\mathbf{k})\rightarrow D_{\alpha}$ with $\alpha=\{e,g\}$ we find
\begin{equation}\label{gk1}
g_{\mathbf{k}}=\Big({\frac{k^2}{2NM\omega_{\mathbf{k}}}\Big)^{1/2}}\sum_{\alpha} D_{\alpha} \int d^3r|\psi_{\alpha}(\mathbf{r})|^2e^{i\mathbf{k}\cdot\mathbf{r}},
\end{equation}
\begin{equation}\label{fkk1}
f_{\mathbf{k,k'}}=\frac{1}{4NM\sqrt{\omega_{\mathbf{k}}\omega_{\mathbf{k'}}}} (k'+k\cos\theta_{kk'})(k+k'\cos\theta_{kk'})\sum_{\alpha} D_{\alpha} \int d^3r |\psi_{\alpha}(\mathbf{r})|^2e^{i(\mathbf{k}+\mathbf{k'})\cdot\mathbf{r}},
\end{equation}
where $\theta_{kk'}$ is the angle between $\mathbf{k}$ and $\mathbf{k'}$ wave vectors.

\section{Master equation}

To calculate the emission spectra, we initialise the system in its excited state, and use a Born-Markov master equation in the polaron frame to calculate the subsequent dynamics. To calculate the intensity correlation function the master equation is derived using the same methodology, though now including a driving term. Here we present the latter derivation applicable to the driven case, and the non-driven case can be reproduced by setting the Rabi frequency $\Omega$ and the laser frequency $\omega_\mathrm{d}$ to zero.

The Hamiltonian describing the laser driven molecule is $H_\mathrm{D}=H_\mathrm{S,D}+H_\mathrm{E}+H_\mathrm{I}^\mathrm{EM}+H_\mathrm{I}^\mathrm{PH-TLS}+H_\mathrm{I}^\mathrm{PH-LV}$, where we have moved into a rotating frame with respect to the driving laser frequency $\omega_\mathrm{d}$. The system Hamiltonian including driving is $H_\mathrm{S,D}=(E_\mathrm{X}-\omega_\mathrm{d}) \sigma^{\dagger}\sigma +\frac{\hbar\Omega}{2}(\sigma+\sigma^{\dagger})+ \hbar\sum_i[\Delta_i a^{\dagger}_ia_i +\eta_i \sigma^{\dagger}\sigma(a^{\dagger}_i+a_i)]$, where the terms are all consistent with those that are defined in the main text. The other term differing to the non-driven Hamiltonian is the electromagnetic (EM)--TLS interaction which picks up a phase in the rotating frame $H_\mathrm{I}^\mathrm{EM}=\hbar\sum_l p_l( e^{-i\hbar\omega_\mathrm{d} t} \sigma c_l^{\dagger}+e^{+i\hbar\omega_\mathrm{d} t} \sigma^{\dagger}c_l)$. The remaining term independent of driving is the phonon bath--local vibrational interaction $H_\mathrm{I}^\mathrm{PH-LV}=\hbar\sum_{i,\mathbf{q}} (h_{i,\mathbf{q}} a_i d_\mathbf{q}^{\dagger}+\mathrm{h.c.})$. The final terms $H_\mathrm{E}$ and $H_\mathrm{I}^\mathrm{PH-LV}$ are fully specified in the main text.

We now perform two polaron transformations on this driven Hamiltonian. The first of these is defined through $H_{P1}=U_{P1} H U_{P1}^{\dagger}$ where $U_{P_1}=\ket{g}\bra{g}+\ket{e}\bra{e}B_+$ with the bath displacement operator $B_{\pm}=\mathrm{exp}[\pm \sum_\mathbf{k}\frac{g_\mathbf{k}}{\omega_\mathbf{k}}(b_\mathbf{k}^{\dagger}-b_\mathbf{k})]$. The second transformation is defined through  $H_{P2}=U_{P2}H_PU^{\dagger}_{P2}$ where $U_{P_2}=\ket{g}\bra{g}+\ket{e}\bra{e}\prod_i \mathcal{B}_i$ with $\mathcal{B}_i=\mathrm{exp}[\frac{\eta_i}{\Delta_i}(a^{\dagger}_i-a_i)]$. After transforming the Hamiltonian into the polaron frame we find 
$H_\mathrm{P,D}=H_0+H_\mathrm{I}$, where $H_0=H_\mathrm{S,D,P}+H_\mathrm{E}$ and 
$H_\mathrm{I}=H_\mathrm{I}^\mathrm{DR}+H_\mathrm{P,I}^\mathrm{EM}+H_\mathrm{I,2}^\mathrm{PH}+H_\mathrm{P,I}^\mathrm{PH-LV},$ with the system term $H_\mathrm{S,D,P}=\delta_\mathrm{P}\sigma^{\dagger}\sigma+\frac{\Omega}{2}\expval{B}\underline{X}+\hbar\sum_i\Delta_i a^{\dagger}_ia_i$. The transformed interaction terms take the form
\begin{equation}
H_\mathrm{I}^\mathrm{DR}=\frac{\hbar\Omega}{2}(\underline{X}B_x+\underline{Y}B_y),
\end{equation}
\begin{equation}\label{HemTLS}
H_\mathrm{P,I}^\mathrm{EM}=\hbar\sum_l p_l e^{-i\hbar\omega_\mathrm{d} t} \sigma_aB_{-} c_l^{\dagger}+\mathrm{h.c.},
\end{equation}
\begin{equation}\label{HPH}
H_\mathrm{P,I}^\mathrm{PH-LV}=\hbar\sum_{i,\mathbf{q}}(a_i^{\dagger}-\frac{\eta_i}{\Delta_i}\sigma^{\dagger}\sigma) h_{i\mathbf{q}}  d_\mathbf{q}+\mathrm{h.c.},
\end{equation}
where $H_\mathrm{I,2}^\mathrm{PH}$ is defined in the main text. 
The system operators above are $\underline{X}=(\sigma_a+\sigma_a^{\dagger})$, $\underline{Y}=i(\sigma_a-\sigma_a^{\dagger})$, with dressed dipole operator $\sigma_a=\sigma \prod_i \mathcal{B}_i$, and bath operators $B_x=\frac{1}{2}(B_++B_--2\expval{B})$ and $B_y=\frac{i}{2}(B_+-B_-)$. We consider the case of resonant driving such that $\delta_\mathrm{P}=E_\mathrm{P}-\hbar\omega_\mathrm{d}=0$, where $E_{\mathrm{P}}=E_\mathrm{X}-\hbar(\sum_i\eta_i^2/\Delta_i+\sum_\mathbf{k}g_\mathbf{k}^2/\omega_\mathbf{k})$ is the  polaron shifted TLS energy splitting.

We now derive the second order Born-Markov master equation 
for the polaron frame reduced density operator in the Schr\"{o}odinger picture $\rho(t)$, which begins from the general form 
\begin{equation}\label{SchME}
\partial_t \rho\mathrm(t) =-\frac{i}{\hbar}[H_\mathrm{S,D,P},\rho\mathrm(t)]-\frac{1}{\hbar^2}\int_0^{\infty} d\tau \Trace_\mathrm{E}\Big( [H_\mathrm{I},[\tilde{H_\mathrm{I}}(-\tau),\rho\mathrm(t) \otimes \rho_\mathrm{E}]] \Big),
\end{equation}
where $\tilde{H}_\mathrm{I}(-\tau)=\mathrm{exp}[-i H_0 \tau/\hbar] H_\mathrm{I} \mathrm{exp}[i H_0 \tau/\hbar]$ is the interaction picture interaction Hamiltonian. The polaron transformed driven Hamiltonian has four interaction terms. We assume that there are no correlations between the phonon and EM environments, and that  fluctuations experienced by each of the local vibrational modes are uncorrelated. With these assumptions cross terms between interaction terms vanish and we can evaluate separately four dissipators corresponding to each of the four interactions terms in the Hamiltonian.

\subsection{Driving dissipator}
We begin with the dissipator arising from the driving-induced interaction Hamiltonian $H_\mathrm{I}^\mathrm{DR}=\frac{\Omega}{2}\underline{X}\otimes B_x +\frac{\Omega}{2}\underline{Y}\otimes B_y$. Moving into the interaction picture for the system operators by using a Fourier decomposition we have $ \underline{X}(-\tau)=\sum_{\xi}e^{i\xi \tau}\underline{X}(\xi)$ and $ \underline{Y}(-\tau)=\sum_{\xi}e^{i\xi \tau}\underline{Y}(\xi)$. The environment operators in the interaction picture are $\tilde{B}_x(-\tau)=e^{-iH_\mathrm{E}\tau/\hbar}B_x e^{iH_\mathrm{E}\tau/\hbar}$ and similarly for $\tilde{B}_y(-\tau)$. We define the environment correlation functions as
\begin{equation}\label{Ecorr}
C_{ij}(\tau)=\Trace_\mathrm{E}(\tilde{B}_i(\tau)B_j\rho_\mathrm{E}(0)),
\end{equation}
and substituting in the relevant operators we find the correlation functions $C_{xx}(\tau)=\frac{\expval{B}^2}{2}(e^{\phi(\tau)}+e^{-\phi(\tau)}-2)$, $C_{yy}(\tau)=\frac{\expval{B}^2}{2}(e^{\phi(\tau)}-e^{-\phi(\tau)})$ and $C_{xy}(\tau)=C_{yx}(\tau)=0$. We then defining the general environment response functions as
\begin{equation}\label{EnvRes}
K_{ij}(\xi)=\int^\infty_0d\tau C_{ij}(\tau) e^{i\xi\tau}=\frac{1}{2}\gamma_{ij}(\xi)+iS_{ij}(\xi).
\end{equation}
Substituting the system operators along with the response functions $K_{xx}(\xi)$ and $K_{yy}(\xi)$ into the Eq.~(\ref{SchME}) we find
\begin{equation}
\begin{split}
\partial_t \rho_\mathrm{S}(t) &=-\frac{i}{\hbar}[H_\mathrm{S}^\mathrm{LS},\rho_\mathrm{S}(t)]-\frac{1}{2}\Big(\frac{\Omega}{2}\Big)^2\sum_{\xi}  \gamma_{xx}(\xi)[\underline{X},\underline{X}(\xi)\rho_\mathrm{S}(t)-\rho_\mathrm{S}(t)\underline{X}^{\dagger}(\xi)]\\&-\frac{1}{2}\Big(\frac{\Omega}{2}\Big)^2\sum_{\xi}  \gamma_{yy}(\xi)[\underline{Y},\underline{Y}(\xi)\rho_\mathrm{S}(t)-\rho_\mathrm{S}(t)\underline{Y}^{\dagger}(\xi)]+\mathcal{D}_\mathrm{x3}[\rho_\mathrm{S}(t)]\\
&=-\frac{i}{\hbar}[H^\mathrm{LS}_\mathrm{S},\rho_\mathrm{S}(t)]+\mathcal{D}_\mathrm{DR}[\rho_\mathrm{S}(t)]+\mathcal{D}_\mathrm{x3}[\rho_\mathrm{S}(t)]
\end{split}
\end{equation}
where $\mathcal{D}_\mathrm{DR}[\tilde{\rho}(t)]$ is the driving dissipator and we have introduced $H^\mathrm{LS}_\mathrm{S}=H_\mathrm{S,D,P}+H_\mathrm{LS}$ which includes a Lamb shift originating from the imaginary component of the response function. The final term $\mathcal{D}_\mathrm{X3}[\rho_\mathrm{S}(t)]$ represents the remaining three dissipator terms.

\subsection{Thermal phonon bath dissipator}

To evaluate the dissipator arising from coupling to the thermal phonon bath, we write Eq.~(\ref{SchME}) instead in the interaction picture: 
\begin{equation}\label{intEM}
\partial_t \tilde{\rho}_\mathrm{S}(t)=-\frac{1}{\hbar^2}\int^{\infty}_0 d\tau \Trace_\mathrm{E} \Big( [\tilde{H}_{I}(t),[\tilde{H}_\mathrm{I}(t-\tau),\tilde{\rho}_\mathrm{S}(t)\rho_\mathrm{E} ]] \Big),
\end{equation}
where $\tilde{\rho}_\mathrm{S}(t) =e^{i H_\mathrm{S,D,P} t/\hbar}\rho_\mathrm{S}(t) e^{-i H_\mathrm{S,D,P} t/\hbar}$. Working in the interaction picture allows the secular approximation to be made which simplifies the algebra. Transforming $H_\mathrm{P,I}^\mathrm{PH-LV}$ into the interaction picture we need to find $\tilde{H}_\mathrm{P,I}^\mathrm{PH-LV}(-\tau)=U_0(-\tau)^{\dagger}H_\mathrm{P,I}^\mathrm{PH-LV} U_0(-\tau)$ where $U_0(-\tau)=e^{i(H_\mathrm{S,P,D}+H_\mathrm{E})\tau/\hbar}$. To proceed we make the approximation 
$e^{-(\frac{\Omega \expval{B}}{2} \underline{X} +\sum_i\Delta_ia_i^{\dagger}a_i)\tau}a_je^{i(\frac{i\Omega \expval{B}}{2}\underline{X}+\sum_i\Delta_ia_i^{\dagger}a_i)\tau}\approx a_je^{-i\Delta_j\tau}$, which is valid as 
$\hbar\Omega\expval{B}(\sim 1\mu \mathrm{eV}) \ll \hbar\Delta_i (20-40~\mathrm{meV})$ for a typical single molecule emitter and using the assumption that local vibrational mode fluctuations are uncorrelated. 
This leads to the an interaction Hamiltonian which does not depend on the driving:
\begin{equation}\label{intHpl}
\tilde{H}_\mathrm{I,P}^\mathrm{PH-LV}(t)=\sum_i(a_ie^{-i\Delta_i\tau}-\frac{\eta_i}{\Delta_i}\sigma^{\dagger}\sigma)\sum_\mathbf{q} h_{i\mathbf{q}}  d^{\dagger}_\mathbf{q}e^{i z_q t}+\mathrm{h.c.}
\end{equation}
Now, by writing $\tilde{H}_\mathrm{I,P}^\mathrm{PH-LV}(t)=\sum_{j=1,2} \tilde{A}_j(t)\tilde{B}_j(t)$ where 
$\tilde{B}_1(t)=\sum_\mathbf{q} h_\mathbf{q}  d_\mathbf{q}e^{-i z_q t}$ and $\tilde{B}_2(t)=\tilde{B}_1^{\dagger}(t)$, we calculate correlation functions according to Eq.~(\ref{Ecorr}), finding 
\begin{equation}\label{corRES1}
C_{12}(\pm \tau)=\int^{\infty}_0 d\nu J_\mathrm{PH-LV}(\nu)n(\nu)e^{\pm i \nu \tau},
\end{equation}
\begin{equation}
C_{21}(\pm \tau)=\int^{\infty}_0 d\nu J_\mathrm{PH-LV}(\nu)(n(\nu)+1)e^{\pm i \nu \tau},
\end{equation}
where we have introduced the local vibrational mode--phonon bath spectral density 
$J_\mathrm{PH-LV}(\nu)=\sum_q |p_q|^2\delta(z_q-\nu)$ and the Bose occupancy number $n(\nu)=(e^{\hbar \nu/k_BT}-1)^{-1}$. 
Inserting (\ref{intHpl}) and the phonon correlation functions into the interaction picture master equation (\ref{intEM}) we find
\begin{equation}\label{intRES2}
\begin{split}
 \partial_t \tilde{\rho}_S(t) &=-\sum_i\int^{\infty}_0 d\tau \Bigg( C_{12}(\tau)[a_i,a_i^{\dagger}\tilde{\rho}_S(t)]e^{-i\Delta_i \tau}+C_{21}(-\tau)[a_i^{\dagger},a_i\tilde{\rho}_S(t)]e^{i\Delta_i \tau}\\
 &+\Big(\frac{\eta_i}{\Delta_i}\Big)^2\big(C_{12}(\tau)+C_{21}(-\tau)\big)[\sigma^{\dagger}\sigma,\sigma^{\dagger}\sigma\tilde{\rho}_S(t)] +\mathrm{h.c.}\Bigg)
\end{split}
\end{equation}
where the secular (rotating wave) approximation has been made. Performing the time integrals and neglecting Lambshift terms that we absorb into our definitions (see Eq.~\ref{EnvRes}) 
we find 
\begin{equation}\label{res12}
\frac{\gamma_{12}(\Delta_i)}{2}=\Re[\int^{\infty}_0 d \tau D_{12}(\tau)e^{-i\Delta_i\tau}]=\pi J_\mathrm{PH-LV}(\Delta_i)n(\Delta_i),
\end{equation}
\begin{equation}\label{res21}
\frac{\gamma_{21}(\Delta_i)}{2}=\pi J_\mathrm{PH-LV}(\Delta_i)(n(\Delta_i)+1).
\end{equation}
When substituting in the super-Ohmic form $J_\mathrm{PH-LV}(\Delta_i)\propto \frac{\Delta_i^3}{\xi^2}e^{-\Delta_i/\xi}$, the final response function $\frac{\gamma_{TLS}(0)}{2}=\Re[ \Big(\frac{\eta_i}{\Delta_i}\Big)^2\int^{\infty}_0 d \tau \big(D_{12}(\tau)+D_{21}(-\tau)\big)]$ is found to be 
$\frac{\gamma_{TLS}(0)}{2}= \Big(\frac{\eta_i}{\Delta_i}\Big)^2 \pi \lim_{\epsilon \rightarrow 0} J_\mathrm{PH-LV}(\epsilon)\coth(\frac{\beta \epsilon}{2})=0$. Substituting these rates back in to equation (\ref{intRES2}) and transforming back into the Schr\"{o}dinger picture, the master equation becomes 
\begin{equation}\label{ME1}
 \partial_t \rho_\mathrm{S}(t) =-i[H_\mathrm{S}^\mathrm{LS},\rho_\mathrm{S}(t)]+\mathcal{D}_\mathrm{PH}[\rho_\mathrm{S}(t)]+\mathcal{D}_\mathrm{DR}[\rho_\mathrm{S}(t)]+\mathcal{D}_{x2}[\rho_\mathrm{S}(t)],
\end{equation}
where $\mathcal{D}_\mathrm{PH}[\rho_\mathrm{S}(t)]=\sum_i\Big(\Gamma_{i,+}\mathcal{L}_{a^{\dagger}}[\rho(t)]+\Gamma_{i,-}\mathcal{L}_{a}[\rho(t)]\Big)$ with Lindblad operator 
$\mathcal{L}_{A}(\rho(t))=A\rho_\mathrm{S}(t)A^{\dagger}- \frac{1}{2}\qty\big{ A^{\dagger}A,\rho_\mathrm{S}(t)}$, and $\Gamma_{i,+}=\kappa_i n(\Delta_i)$ and $\Gamma_{i,-}=\kappa_i( n(\Delta_i)+1)$ with  $\kappa_i=\pi J_\mathrm{PH-LV}(\Delta_i)$. The remaining two dissipators are represented by the term $\mathcal{D}_{x2}[\rho_\mathrm{S}(t)]$.

\subsection{Spontaneous emission dissipator}
To find the TLS--EM field dissipator the master equation is evaluated in the Schr\"{o}dinger picture, see Eq.~(\ref{SchME}). 
We first find the interaction picture interaction Hamiltonian  $\tilde{H}_\mathrm{P,I}^\mathrm{EM}(-\tau)=U_0(-\tau)^{\dagger}H_\mathrm{P,I}^\mathrm{EM} U_0(-\tau)$, which we write 
\begin{equation}\label{HintEM}
\tilde{H}_\mathrm{P,I}^\mathrm{EM}(-\tau)=\sigma_a(-\tau)  e^{-i\omega_\mathrm{d} t} B_{-}(-\tau)C^{\dagger}(-\tau)+\mathrm{h.c.},
\end{equation}
where $B_{\pm}(-\tau)=e^{\pm\sum_ k \frac{g_\mathbf{k}}{\omega_\mathbf{k}} (b^{\dagger}_\mathbf{k} e^{-i \omega_\mathbf{k} \tau}-b_\mathbf{k} e^{i \omega_\mathbf{k} \tau})}$  and $C(-\tau)=\hbar\sum_l p_l c_l e^{i\omega_l \tau}$. 
For the system operators in the interaction picture we have
\begin{equation}
\begin{split}
 \sigma_a(-\tau) e^{-i\omega_\mathrm{d} t}&=e^{-i(\frac{\Omega \expval{B}}{2} \underline{X} +\sum_i\Delta_i a_i^{\dagger}a_i)\tau}\sigma_a e^{i(\frac{i\Omega \expval{B}}{2} \underline{X} +\sum_i\Delta_i a_i^{\dagger}a_i)\tau} e^{-i\omega_\mathrm{d} t},\\
&\approx \sigma  e^{-\sum_i\frac{\eta_i}{\Delta_i}(a_i^{\dagger}e^{-i\Delta \tau}-a_i e^{i\Delta_i \tau})} e^{-iE_\mathrm{P} t},
\end{split}
\end{equation}
where we have again used the large difference in energy scales $E_\mathrm{P}\sim\omega_\mathrm{d}\sim 1.5\,\mathrm{eV}$ compared to $\Omega\expval{B}\approx 10\,\mathrm{meV}$. 
Similarly to the phonon dissipator, with this approximation this is equivalent to the non-driven electromagnetic dissipator detailed in the main text. 

From Eq.~({\ref{HintEM}}), we see the correlation functions for this dissipator have contributions from both the electromagnetic and thermal phonon environments. The electromagnetic environment correlation function is $\mathrm{Tr}_E(CC^{\dagger}(-\tau)\rho_\mathrm{E})=\int^\infty_0 d\omega J_\mathrm{EM}(\omega)e^{i\omega \tau}=\chi(\omega)$, where $J_\mathrm{EM}(\omega)$ is the electromagnetic spectral density. The phonon correlation function shown in the main text is $\mathcal{G}(\tau)= e^{\phi(\tau)}\expval{B}^2$. To evaluate the response functions, 
we write $\sigma_a(-\tau)=\sigma e^{-i E_\mathrm{P}\tau}\mathcal{A}(-\frac{\eta}{\Delta},-\tau)$ where, $\mathcal{A}(-\frac{\eta}{\Delta},-\tau)=e^{-\frac{\eta}{\Delta}(a^{\dagger}e^{-i\Delta \tau}-a e^{i\Delta \tau})}$ is the time evolved displacement operator. Expressing this time evolved operator in terms of system Hamiltonian unitary operator $U_S(-\tau)=e^{i\sum_i\Delta_i a_i^{\dagger}a_i \tau}$, we find $\mathcal{A}(-\frac{\eta}{\Delta},-\tau)=U_S(-\tau)\mathcal{A}(-\frac{\eta}{\Delta})U_S^{\dagger}(-\tau)=\sum_{n,m}\bra{n}\mathcal{A}(-\frac{\eta}{\Delta})\ket{m}\ket{n}\bra{m}e^{-i\Delta(m-n)\tau}$. The operator $\sigma_a(-\tau)$ can therefore be expressed as
\begin{equation}\label{siga}
\sigma_a(-\tau)=\sigma \sum_{n,m}\mathcal{A}_{n,m}(-\frac{\eta}{\Delta})e^{-i(E_\mathrm{P}+\hbar\Delta(m-n))\tau/\hbar},
\end{equation}
where $\mathcal{A}_{n,m}(\pm \frac{\eta}{\Delta})=\sum_{n,m}\bra{n}\mathcal{A}(\pm \frac{\eta}{\Delta})\ket{m}\ket{n}\bra{m}$ are the matrix elements of the system displacement operator. 

The spontaneous emission rate is the real part of the response function, which we find to be
\begin{equation}
\begin{split}
\frac{\gamma(\epsilon_{n,m})}{2}&=\Re\Big[\int^{\infty}_0 d\tau \mathcal{G}(0)\chi(\tau)e^{-i\epsilon_{n,m}\tau/\hbar}\Big],\\
&=\Re\Big[\int^{\infty}_0 d\omega \int^{\infty}_0 d\tau J_\mathrm{EM}(\omega) e^{i\big(\hbar\omega-\epsilon_{n,m})\big) \tau/\hbar}\Big]=\pi J_\mathrm{EM}(\epsilon_{n,m}).
\end{split}
\end{equation}
where $\epsilon_{n,m}=E_\mathrm{P}+\hbar\Delta(m-n)$. The spectral density is approximated to be flat over the relevant frequency scales with respect to the molecule emitter, such that $\mathcal{G}(\tau) \rightarrow \mathcal{G}(0)$ and $J_\mathrm{EM}(\epsilon_{n,m})\approx \Gamma_1/\pi$~\cite{Dara_2013}. Inserting $\tilde{H}_\mathrm{P,I}^\mathrm{EM}(-\tau)$, $H_\mathrm{P,I}^\mathrm{EM}$ and substituting the emission rate in to the master equation (\ref{SchME}) we then find 
\begin{equation}
\partial_t \rho_\mathrm{S}(t)=-\frac{i}{\hbar}[H_\mathrm{S}^{LS},\rho_\mathrm{S}(t)]+\Gamma_1\mathcal{L}_{\sigma_a}[\rho(t)]+\mathcal{D}_{PH}[\rho_\mathrm{S}(t)]+\mathcal{D}_{DR}[\rho_\mathrm{S}(t)]+\mathcal{D}_{PD}[\rho_\mathrm{S}(t)],
\end{equation}
where $\mathcal{D}_{PD}[\rho_\mathrm{S}(t)]$ is the pure dephasing is the final dissipator.

\subsection{Pure dephasing dissipator}

The pure dephasing dissipator originates from the quadratic electron--phonon interaction term. In the interaction picture we have $\tilde{H}_\mathrm{I,2}^\mathrm{PH}(-\tau)=U_0(-\tau)^{\dagger}H_\mathrm{I,2}^\mathrm{PH} U_0(-\tau)$ giving
\begin{equation}
\tilde{H}_\mathrm{I,2}^\mathrm{PH}(-\tau)=\sigma^{\dagger}\sigma\sum_{\mathbf{kk'}} f_{\mathbf{kk'}}B_\mathbf{k}(-\tau)B_{\mathbf{k'}}(-\tau),
\end{equation}
where $B_\mathbf{k}(-\tau)=( b_{\mathbf{k}}^{\dagger}e^{-i\hbar\omega_\mathbf{k}\tau}+b_{\mathbf{k}}e^{i\hbar\omega_\mathbf{k}\tau})$. Inserting $H_\mathrm{I,2}^\mathrm{PH}$ and $\tilde{H}_\mathrm{I,2}^\mathrm{PH}(-\tau)$ into Eq.~(\ref{SchME}) 
we find the dissipator takes the form
\begin{equation}
\mathcal{D}_\mathrm{PD}[\rho_\mathrm{S}(t)]=2\gamma \mathcal{L}_{\sigma^{\dagger}\sigma}[\rho(t)],
\end{equation}
where the pure dephasing rate is
\begin{equation}
\gamma=\Re\Big[\int^t_0 d\tau \sum_{\mathbf{kk'}}\abs{f_{\mathbf{kk'}}}^2 \expval{B_\mathbf{k}B_\mathbf{k}(-\tau)}\expval{B_{\mathbf{k'}}B_{\mathbf{k'}}(-\tau)}\Big].
\end{equation}
The factorisation of the correlation function above has been made based on the assumption that phonons do not scatter into the same mode i.e. $\mathbf{k}\ne \mathbf{k'}$ \cite{Reigue2017}. To evaluate the environment correlation functions we move in to the continuum limit $\sum_{\mathbf{kk'}}\rightarrow \frac{V^2}{(2\pi)^6}\int^{\infty}_0d^3k\int^{\infty}_0d^3k' $ 
which gives 
\begin{equation}\label{intg}
\gamma=\frac{V^2}{(2\pi)^6} \int^{\infty}_0 d^3k\int^{\infty}_0 d^3k' \abs{f_{\mathbf{kk'}}}^2 \big(n(\omega_\mathbf{k})(n(\omega_{\mathbf{k}'})+1)\delta(\omega_\mathbf{k}-\omega_\mathbf{k}')+(n(\omega_\mathbf{k})+1)n(\omega_{\mathbf{k}'})\delta(\omega_{\mathbf{k}'}-\omega_{\mathbf{k}}).
\end{equation}
Where we have used the definition $\delta(x-a)=\frac{1}{\pi}\Re \big[\int^{\infty}_0 d\tau e^{i(x-a)\tau}\big]$ and assumed linear dispersion $\omega_{\mathbf{k}}=c|\mathbf{k}|$ where $|\mathbf{k}|=k$ and $c$ is the speed of sound in the nano-crystal. Using the delta functions leads to non-zero values of the integral for the case $k=k'$. 
The coupling constant $\abs{f_{\mathbf{kk}'}}^2$ therefore only needs to be evaluated for $k=k'$. 
Substituting in an isotropic Gaussian function 
$\psi_{\alpha}(\mathbf{r})=(d_{\alpha}\sqrt{\pi})^{-3/2}e^{-r^2/2 d_{\alpha}^2}$
where $d_{\alpha}$ is the confinement potential for the ground and excited states which is assumed to be equal, such that $d_{\alpha} \rightarrow d$ We then find 
\begin{equation}
\abs{f_{\mathbf{kk'}}}^2=\Big(\frac{k}{4NMc}\Big)^2(1+\cos\theta)^4\sum_{\alpha} D^2_{\alpha} e^{-k^2(1+\cos\theta)d^2},
\end{equation}
where we have written $\mathbf{k}\cdot \mathbf{k'}=kk'\cos\theta$. 
Substituting this quadratic coupling constant and converting variables from wavevector magnitude into frequency, as well as defining the phonon cut off frequency $\omega_c=\sqrt{2}c/d$, we find the pure dephasing rate
\begin{equation}
\gamma=\frac{V^2}{128\pi^3(NM)^2c^8} \int^{\infty}_0 d\omega  \omega^6 n(\omega)(n(\omega)+1) \sum_{\alpha} D^2_{\alpha} \int^{\pi}_{0} d\theta \sin(\theta)(1+\cos(\theta))^4 e^{-2\omega^2(1+\cos(\theta) )/\omega_c^2}.
\end{equation}

\begin{figure}[t!]
\centering
  \includegraphics[scale=2]{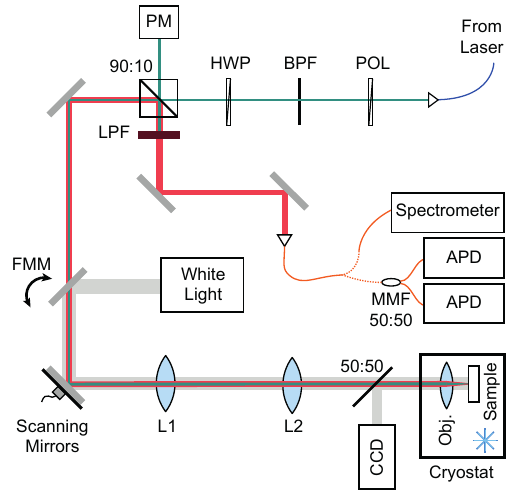}
\caption{Schematic diagram of the confocal microscope. Dark green beam indicates the pump light, red is the fluorescence and grey is the white light used for imaging. Pol: polarizer; BPF: band-pass filter; HWP: half-wave plate; 90:10: 90\% reflection, 10\% transmission cube beam splitter; PM: power meter;  FMM: flip-mount mirror; L1: first lens; L2: second lens; 50:50: 50\% reflection, 50\% transmission pellicle beam splitter Obj.: microscope objective lens; CCD: charge-coupled device camera;  LPF: long-pass filter; MMF 50:50: 50\% reflection, 50\% transmission multimode fibre beam splitter; APD: avalanche photodiode.}
\label{confocal}
\end{figure}

\section{Experimental details}

\begin{figure}[t!]
\centering
  \includegraphics[scale=0.9]{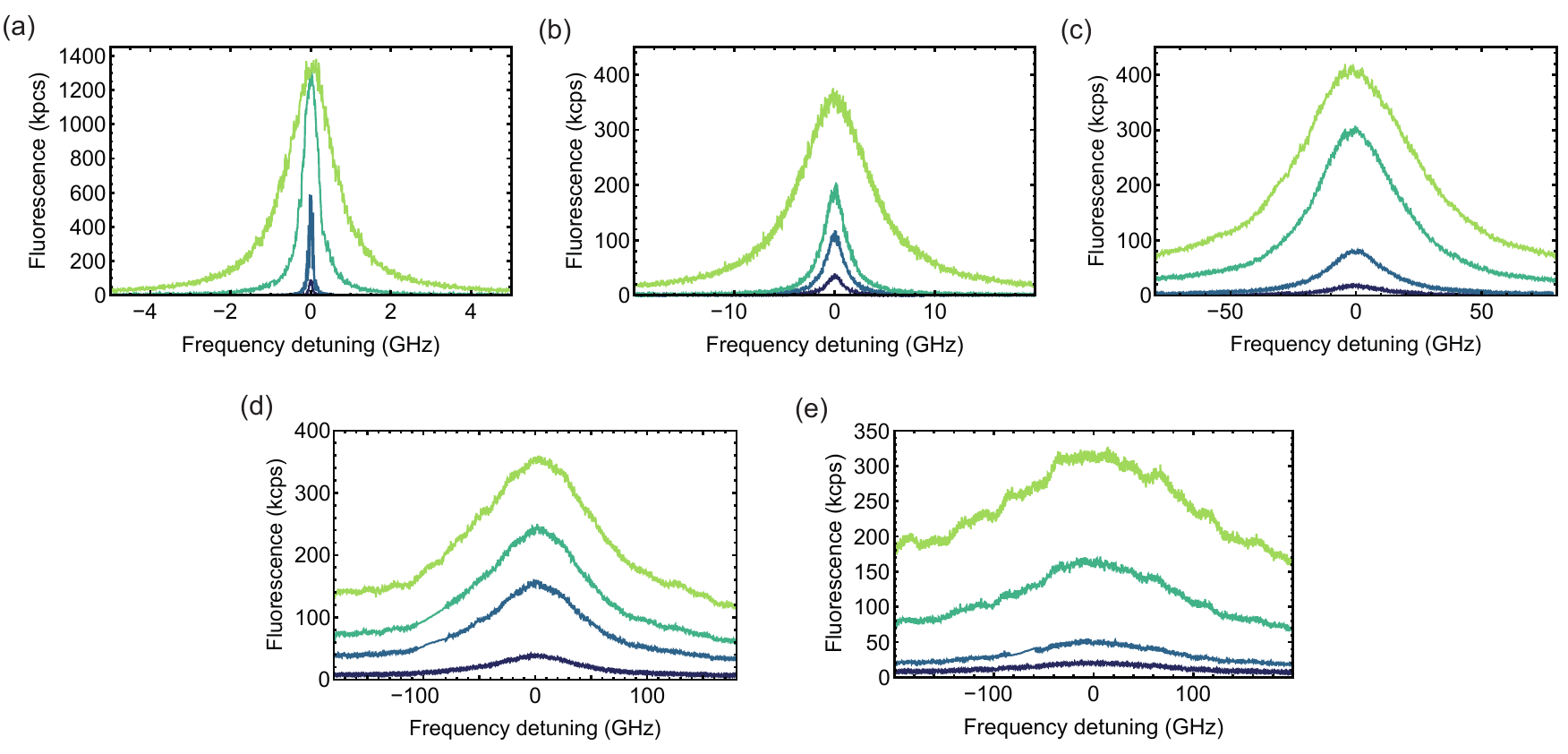}
\caption{Detected red-shifted fluorescence as a narrow laser is tuned across the ZPL of a single DBT molecule for varying illumination intensity at a temperature of (a) 4.7\,K, (b) 10\,K, (c) 20\,K, (d) 31\,K, and (e) 40\,K. All show power broadening and saturation.}
\label{powerbroaden}
\end{figure}

A DBT containing anthracene nanocrystal solution was grown using a recently developed re-precipitation technique \cite{Pazzagli2018}. 5\,$\mu$Mol of 1\,$\mu$l DBT in toluene (VWR) solution was added to 10\,ml of 5\,mMol zone-refined anthracene (Tokyo Chemical Industry UK) in acetone (VWR) solution. 250\,$\mu$l of this mixed solution was then added to 5\,ml of distilled water and sonicated at 37\,kHz for 30~minutes. This solution was filtered through a 450\,nm pore size syringe filter (Sartorius Minisart) and a 25\,$\mu$l drop was then pipetted onto a pre-prepared substrate and left to dry through evaporation. The substrate was a silica-on-silicon wafer that had a 150\,nm layer of gold deposited on the surface to increase collected emission, with a 220\,nm TiO$_2$ spacer layer to protect against plasmonic losses. A protective 150\,nm poly-vinyl alcohol (99~\%+ hydrolyzed, Aldrich) polymer layer was then spin-coated on top and the sample was cooled down to 4.7\,K in a closed-cycle cryostat (Montana Cryostation).

\begin{figure}[t!]
\centering
  \includegraphics[scale=1]{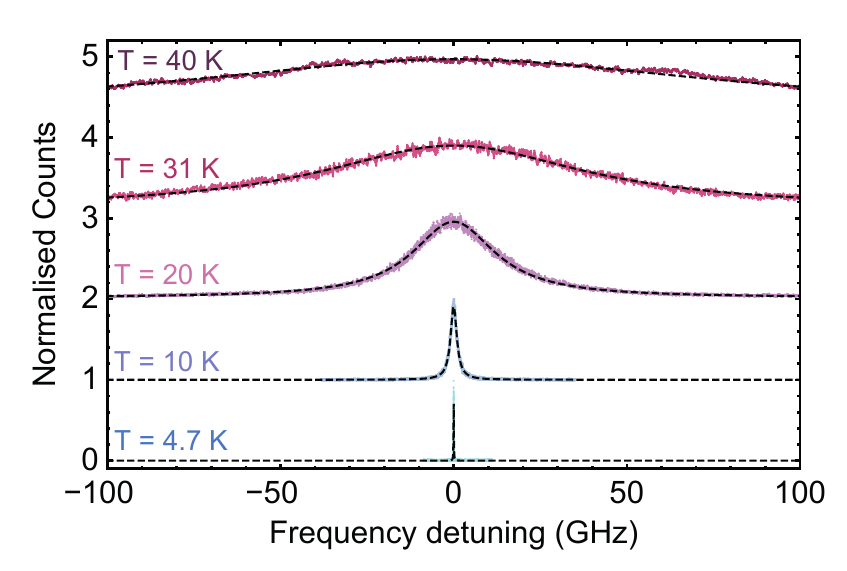}
\caption{Low-power resonant line scans of the ZPL of a single molecule for increasing temperature, showing the temperature-induced broadening of the ZPL. All data is normalised to the peak, and the data for varying temperature has been offset for clarity. Dashed black lines are Lorentzian fits to the data.}
\label{tempbroaden}
\end{figure}

We used a confocal microscope, shown in Fig.~\ref{confocal}, to identify a spatially and spectrally isolated molecule. The same single molecule was used to take all of the data presented here and in the main manuscript. Excitation was performed using a Ti:Sapphire laser (MSquared SolsTIS). The zero-phonon-line wavelength of 782.32\,nm was found by scanning the laser wavelength across the transition and recording the red-shifted fluorescence reaching our APD, whilst rejecting the laser and resonant emission light using an 800\,nm long-pass filter. A Lorentzian line profile was also fitted to this data to determine the linewidth. This process was carried out with increasing illumination power to measure power broadening and saturation \cite{Grandi2016}. Scans at various temperatures for increasing illumination intensity are shown in Fig.~\ref{powerbroaden}. In all cases there is clear power broadening of the linewidth. Low power line scans for the five temperatures investigated are shown in Fig.~\ref{tempbroaden}, with fitted Lorentzian lines.

The second-order correlation function of the emitted light $g^{(2)}(\tau)$ was measured by exciting resonantly and measuring coincidences between red-shifted photons sent to a Hanbury Brown-Twiss interferometer, using two silicon avalanche photodiode (APD) single photon detectors and a timing unit (Picoquant Hydraharp). This allowed us to verify a single molecule was being measured.

\begin{figure}[b!]
\centering
  \includegraphics[scale=1]{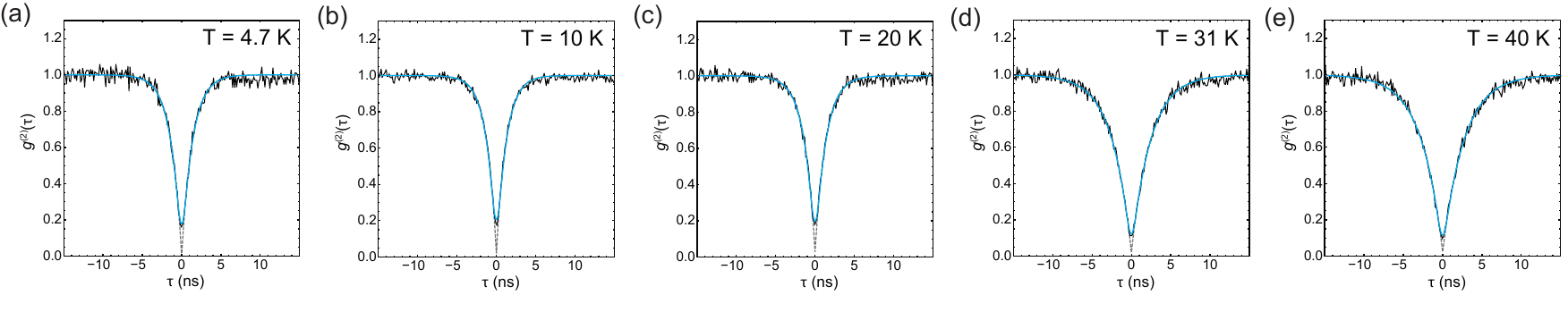}
\caption{Second-order correlation function $g^{(2)}(\tau)$ measurement of light emitted by a DBT molecule pumped to a higher vibrational level of the excited state with 764\,nm light at temperatures of (a) 4.7\,K, (b) 10\,K, (c) 20\,K, (d) 31\,K, and (e) 40\,K. Black lines are the data, blue solid lines a fit including convolution with the finite detector timing jitter, and gray dashed lines the deconvolved case. In all plots $g^{(2)}(0)<0.5$, showing we are collecting light from a single molecule.}
\label{blueg2}
\end{figure}

The spectrum was measured by tuning the laser to 764\,nm and driving the molecule to an excited vibrational level of the excited electronic state. The molecule then undergoes a fast non-radiative decay process to the ground vibrational level of the excited electronic state, from which it decays emitting a photon. The collected light was filtered with a 780\,nm long-pass filter to remove the residual pump laser. The remaining fluorescence, both resonant and red-shifted, was sent to our spectrometer (Andor Shamrock 303i) where it was dispersed by a grating and detected on an EMCCD (Andor Newton). A reference spectrum for background subtraction was taken by spatially moving the beam away from the molecule and repeating the process. This series of measurements was repeated across a range of temperatures between 4.7\,K and 40\,K. Again, to ensure the signal was only originating from a single molecule we measured the second-order correlation function $g^{(2)}(\tau)$ of the light, this time generated via non-resonant excitation for each temperature. The results are shown in Fig.~\ref{blueg2}, and confirmed that for all temperatures that $g^{(2)}(0)<0.5$. Fits to the data are of the form \cite{Grandi2016}
\begin{equation}
    g^{(2)}(\tau) = 1 - V e^{-(1+S)\Gamma_1|\tau|} \, ,
\end{equation}
where $S$ is the saturation parameter and $V$ is a visibility term accounting for background from the laser or other molecules. 

Additionally, a pulsed Ti:Sapphire laser (Spectra Physics Tsunami) was used to resonantly excite the molecule and measure its excited state lifetime, and therefore $\Gamma_1$, by monitoring the time difference between the laser pulse and the detection of a photon. Using this method we find a lifetime of 4.31(3)\,ns, giving $\Gamma_1 = 0.231(2)\,\mathrm{ns}^{-1}$, for the single DBT molecule used throughout this work.

\end{document}